\documentclass[reprint,superscriptaddress,showpacs,amsmath,amssymb,prl,aps,twocolumn]{revtex4-1}

\usepackage{amsmath}
\usepackage{amssymb}
\usepackage{amsfonts}
\usepackage{graphics}
\usepackage{graphicx}
\usepackage{subfigure}
\usepackage{dcolumn}

\usepackage[pdfpagemode=UseNone,pdfstartview=FitH,colorlinks=true,linkcolor=blue,urlcolor=blue,anchorcolor=blue,citecolor=blue]{hyperref}
\allowdisplaybreaks[4]

\begin{document}

\title{Nonreciprocity and Unidirectional Invisibility in Cavity Magnonics}

\author{Yi-Pu Wang}
\email{Email: Yipu.Wang@umanitoba.ca}
\affiliation{Department of Physics and Astronomy, University of Manitoba, Winnipeg, Canada R3T 2N2}

\author{J.W. Rao}
\affiliation{Department of Physics and Astronomy, University of Manitoba, Winnipeg, Canada R3T 2N2}

\author{Y. Yang}
\affiliation{Department of Physics and Astronomy, University of Manitoba, Winnipeg, Canada R3T 2N2}

\author{Peng-Chao Xu}
\affiliation{Department of Physics and Astronomy, University of Manitoba, Winnipeg, Canada R3T 2N2}
\affiliation{State Key Laboratory of Surface Physics and Department of Physics, Fudan University, Shanghai 200433, China}

\author{Y.S. Gui}
\affiliation{Department of Physics and Astronomy, University of Manitoba, Winnipeg, Canada R3T 2N2}

\author{B.M. Yao}
\affiliation{State Key Laboratory of Infrared Physics, Chinese Academy of Sciences, Shanghai 200083, People's Republic of China}

\author{J. Q. You}
\affiliation{Interdisciplinary Center of Quantum Information and Zhejiang Province Key Laboratory of Quantum Technology and Device, Department of Physics and State Key Laboratory of Modern Optical Instrumentation, Zhejiang University, Hangzhou, China}

\author{C.-M. Hu}
\email{Email: hu@physics.umanitoba.ca}
\affiliation{Department of Physics and Astronomy, University of Manitoba, Winnipeg, Canada R3T 2N2}

\date{\today}
\begin{abstract}
We reveal the cooperative effect of coherent and dissipative magnon-photon couplings in an open cavity magnonic system, which leads to nonreciprocity with a considerably large isolation ratio and flexible controllability. Furthermore, we discover unidirectional invisibility for microwave propagation, which appears at the zero-damping condition for hybrid magnon-photon modes. A simple model is developed to capture the generic physics of the interference between coherent and dissipative couplings, which accurately reproduces the observations over a broad range of parameters. This general scheme could inspire methods to achieve nonreciprocity in other systems.

\end{abstract}
\maketitle

\textit{Introduction.-} Reciprocity is ubiquitous in nature, but sensitive signal detection and processing, especially in the delicate quantum regime~\cite{Walls-94}, requires nonreciprocal operations. The case of the greatest interest is electromagnetic nonreciprocity~\cite{Caloz-18} which focuses on the nonreciprocal propagation of electromagnetic fields ranging from microwave~\cite{Chapman-17,Bernier-17,Peterson-17,Lecocq-17}, terahertz~\cite{Shalaby-13}, optical~\cite{Miri-17,Sounas-2017,Ramezani-18,Fang-17,Shen-16} to X-ray frequencies~\cite{Goulon-2000}. In addition, acoustic \cite{Fleury-14,Walker-18} and  phononic \cite{yifan-18,Torrent-18} nonreciprocities have been reported and have their own applications. Nonreciprocity may be achieved using different mechanisms, such as Faraday rotation~\cite{Faraday,Hogan-53, Rowen-53,Adam-2002,Camley-09}, optomechanical interactions~\cite{Peterson-17,Shen-16,Miri-17,Bernier-17}, reservoir engineering~\cite{Clerk-15,Fang-17}, and parametric time modulation~\cite{Ranzani-14,Sounas-2017}. However, in practice, achieving large nonreciprocity with flexible controllability remains a demanding issue, especially in the linear response regime~\cite{Caloz-18,Fan-15}.

Recently, cavity magnonics has attracted a lot of attention~\cite{Huebl-13,Tabuchi-14,Zhang-14,Tobar-14,Hu-15,Cao-15,Hu-17,Dengke-17}. Such hybrid systems show great application prospects in quantum information processing, acting either as a quantum transducer~\cite{TabuchiScience-15,NakamuraSA-17,Osada-16,Zhangxu-16,haighprl-16,Hisatomi-16,Braggio-17} or quantum memory~\cite{TangNC-15}. A versatile magnon-based quantum information processing platform has taken shape \cite{Dany-19}. However, in conventional cavity magnonics systems governed by coherent magnon-photon coupling, nonreciprocity is missing \cite{Harder-18s}. Last year, an intriguing dissipative magnon-photon coupling was discovered \cite{Harder-18,Xia-18}, which has quickly been verified as ubiquitous \cite{Bhoi-19,Ying-19,Rao-19,Boventer-19,Igor-19,Yao-19,Yu-19}. Microscopically, dissipative coupling results from the traveling-wave-induced \cite{Yao-19} cooperative external coupling \cite{Clerk-15,Yu-19,Toth-17}. The interference of coherent and dissipative magnon-photon couplings makes it possible to break time-reversal symmetry, which we demonstrate in this letter. Furthermore, we show that such a unique scheme of nonreciprocity has flexible controllability, which enables high isolation ratio accompanied by low insertion loss. Our results provide a priori reference for exploring nonreciprocity at different frequencies based on the interference of coherent and dissipative couplings in other systems.

\begin{figure}[!t]
  \centering
  \includegraphics[width=0.4\textwidth]{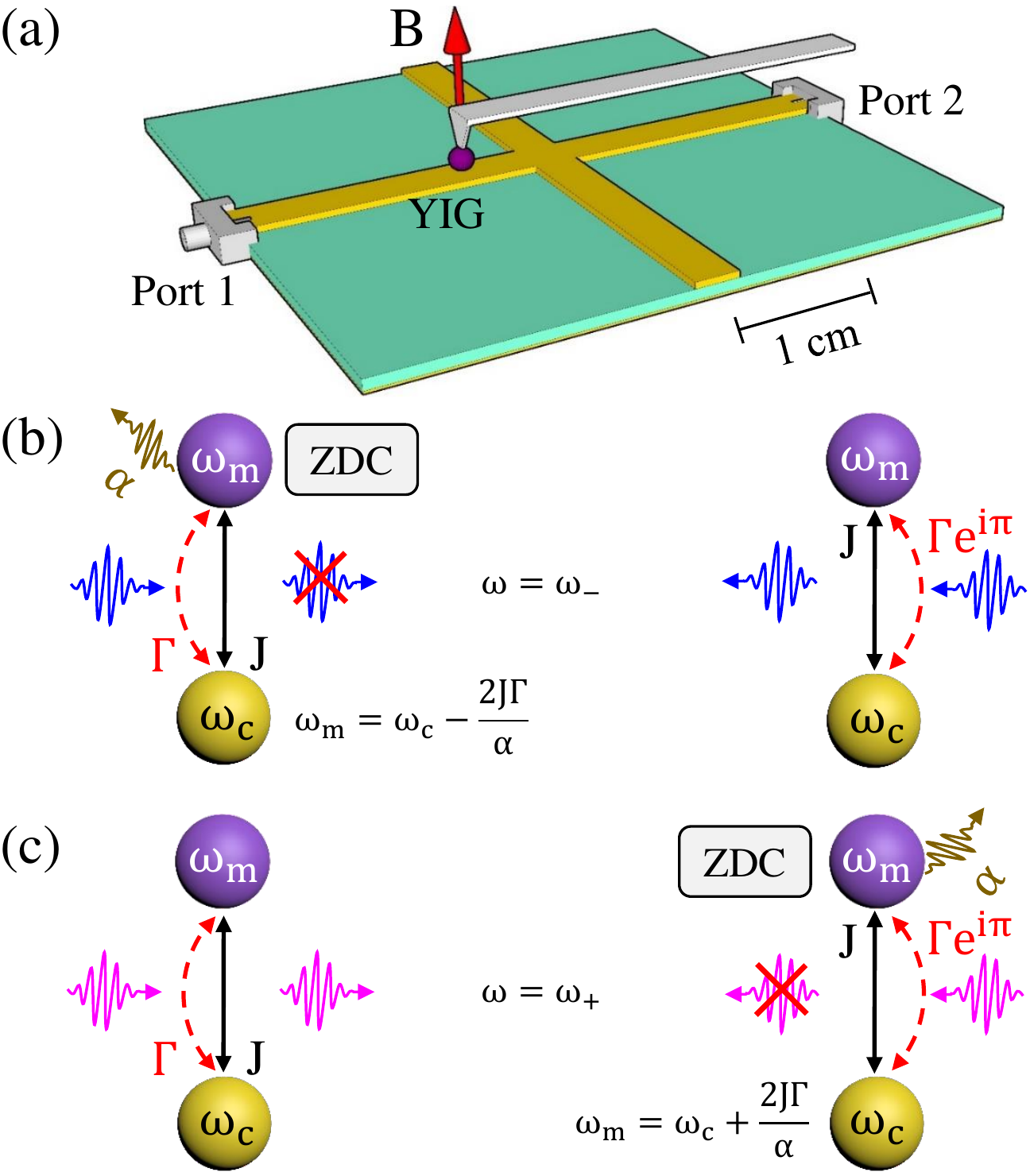}
  \caption{(color online). (a)~Schematic diagram of the experimental setup. The cross-line cavity supports both standing cavity modes and traveling waves. A YIG sphere is glued on the end of a displacement cantilever. The cantilever can be adjusted in 3D space precisely. The spectroscopy is measured with a vector network analyzer through port 1 and 2. (b)(c)~Schematic diagram showing magnon-photon coupling and nonreciprocal microwave transmission at the zero-damping conditions (ZDCs).}
  \label{fig:1}
\end{figure}

\textit{System and model.-}The experimental setup is schematically shown in Fig.~\ref{fig:1}(a), where a yttrium iron garnet (YIG) sphere (1-mm diameter) is placed close to a cross-line microwave circuit. The circuit is designed to support both standing and traveling waves~\cite{supp}, forming an open cavity where the external coupling is much larger than the intrinsic damping rate, and it may even become comparable to the cavity eigenfrequency. The YIG sphere is glued on the end of a displacement cantilever, enabling 3D-position control. A magnetic field of a few hundred millitesla is applied perpendicular to the cavity plane, controlling the magnon mode frequency $\omega_{\rm{m}}$. The cavity mode at the frequency $\omega_{\rm{c}}$ contributes to the coherent magnon-photon coupling (with the coupling rate $J$). The traveling wave causes the dissipation of magnons through radiation damping to the environment~\cite{Yao-19}, inducing an effective dissipative coupling (with the rate $\Gamma$) between the cavity and the magnon modes~\cite{Clerk-15,Yu-19,supp}. In our setup, both $J$ and $\Gamma$ can be drastically changed when moving the YIG sphere. Furthermore, according to Amp\'ere's law, the phase of the rf field of the traveling wave at the YIG position differs by $\pi$ by reversing the propagation direction. Hence, the relative phase $\Theta$ between the coherent and dissipative couplings differs by $\pi$ for microwave signals loaded from port 1 and port 2. As we will show below, these ingredients, i.e., the cooperation of coherent and dissipative couplings, and their direction-dependent relative phase, govern the physics for the new scheme of nonreciprocity.    

To describe such an open cavity magnonic system, we construct the following non-Hermitian Hamiltonian that takes into account the loading configurations~\cite{supp}:
\begin{equation}\label{Hamiltonian}
\begin{split}
\hat{H}/\hbar=&\widetilde{\omega}_{\rm{c}}\hat{a}^{\dag}\hat{a}+\widetilde{\omega}_{\rm{m}}\hat{b}^{\dag}\hat{b}+(J-i\Gamma e^{i\Theta})(\hat{a}^{\dag}\hat{b}+\hat{b}\hat{a}^{\dag}).\\
\end{split}
\end{equation}
Here, $\widetilde{\omega}_{\rm{c}}=\omega_{\rm{c}}-i\beta$, $\widetilde{\omega}_{\rm{m}}=\omega_{\rm{m}}-i\alpha$, $\hat{a} (\hat{a}^\dag)$ and $\hat{b} (\hat{b}^\dag)$ are the photon and magnon annihilation (creation) operators, respectively, and $\alpha$ and $\beta$ are the intrinsic damping rates of the magnon and cavity modes, respectively. Without losing generality, we choose $\Theta$ = 0 and $\pi$ for microwaves loaded from port 1 and 2, respectively. 

The eigenvalues of Eq.~(\ref{Hamiltonian}), $\widetilde{\omega}_{\pm}=\big[\widetilde{\omega}_{\rm{c}}+\widetilde{\omega}_{\rm{m}}\pm\sqrt{(\widetilde{\omega}_{\rm{c}}-\widetilde{\omega}_{\rm{m}})^2+4(J-ie^{i\Theta}\Gamma)^2}\big]/2$, correspond to two hybridized modes that have intriguing properties. In particular, their intrinsic damping rate Im($\widetilde{\omega}_{\pm}$) may go to zero, which we define as the zero-damping condition (ZDC)~\cite{supp}. For example, when $J$ = $\Gamma >\alpha$, $\beta$, and for $\Theta$ = 0, ZDCs appear when $\omega_{\rm{m}}=\omega_{\rm{c}}-2J\Gamma /\alpha$ and $\omega_{\rm{m}}\approx\omega_{\rm{c}}+2J\Gamma /\beta$. For $\Theta$ = $\pi$, the ZDCs appear when $\omega_{\rm{m}}=\omega_{\rm{c}}+2J\Gamma /\alpha$ and $\omega_{\rm{m}}\approx\omega_{\rm{c}}-2J\Gamma /\beta$. Peculiar transmission properties emerge at ZDCs, which we discover by calculating the transmission coefficient of the system using input-output theory~\cite{Walls-94,supp}, 
\begin{eqnarray}\label{J-s12}
S_{21(12)}=1+\frac{\kappa}{i(\omega-\omega_{\rm{c}})-(\kappa+\beta)+\frac{-[iJ+\Gamma e^{i\Theta_{1(2)}}]^{2}}{i(\omega-\omega_{\rm{m}})-(\alpha+\gamma)}}.
\label{J-s12}
\end{eqnarray}
Here, $\Theta_{1}$=0, $\Theta_{2}$=$\pi$, and $\kappa$ and $\gamma$ are the external damping rates~\cite{supp} of the cavity and magnon modes, respectively. 

\begin{figure}[t]
	\centering
	\includegraphics[width=0.47\textwidth]{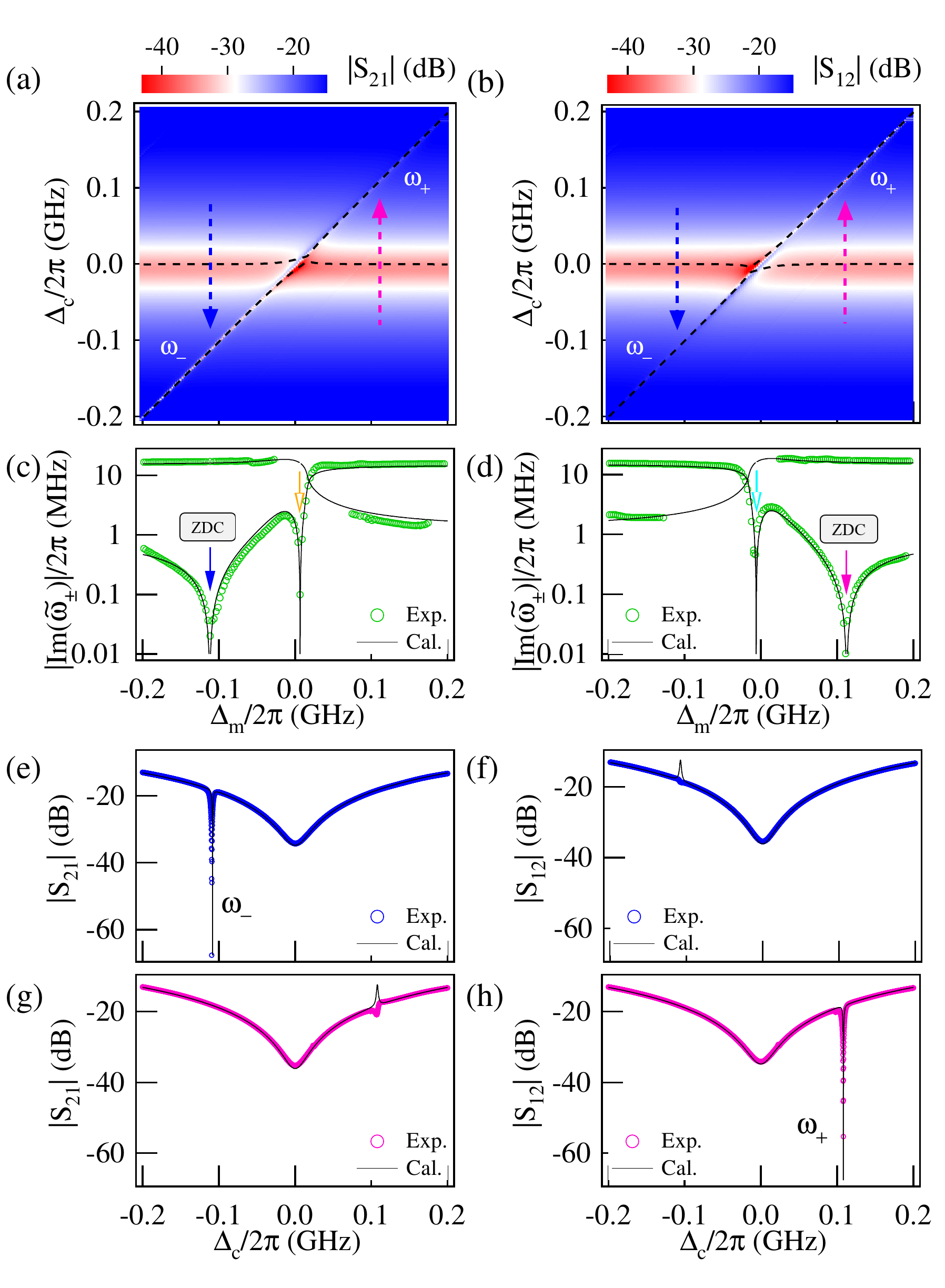}
	\caption{(color online). Mapping of (a) $|S_{21}|$ and (b) $|S_{12}|$ spectra measured when $J/2\pi=\Gamma/2\pi$ =7.9 MHz as a function of $\Delta_{\rm{c}}$ and $\Delta_{\rm{m}}$.  Blue and magenta arrows mark the bias fields at which the spectra plotted in (e)-(h) are measured. The black dashed lines are the calculated dispersion curves. (c)(d)~Green circles are extracted intrinsic damping rates of the hybridized modes as a function of $\Delta_{\rm{m}}$. The black solid lines are the calculated imaginary part of the eigenvalues. The zero-damping conditions are marked by arrows.  (e)~$S_{21}$ and (f) $S_{12}$ spectra measured at the field marked by the blue arrow. (g)~$S_{21}$ and (h) $S_{12}$ spectra measured at the field marked by the magenta arrow. Thin curves in (e)-(h) are calculated from Eq. (2).}
	\label{fig:2}
\end{figure}

Noticeably, a term of $-2iJ\Gamma e^{i\Theta_{1(2)}}$ arises in Eq.~(\ref{J-s12}) due to the interference of coherent and dissipative couplings. This term produces nonreciprocity. Most remarkably, we find that~\cite{supp} for ZDCs at $\omega_{\rm{m}}=\omega_{\rm{c}}\pm2J\Gamma /\alpha$:
\begin{subequations}
\begin{align}
|S_{21} (\omega_{-})| & = |S_{12} (\omega_{+})| = 0, \\
|S_{12} (\omega_{-})| & = |S_{21} (\omega_{+})| > 0,
\end{align}
\label{Symmetry}
\end{subequations}
where $\omega_{\pm}$ = Re($\widetilde{\omega}_{\pm}$) are the frequencies of the hybridized modes, and $S_{21(12)}$ is in linear scale. Similar symmetry applies for the other pair of ZDCs at $\omega_{\rm{m}}\approx\omega_{\rm{c}}\pm2J\Gamma /\beta$. Eq.~(\ref{Symmetry}a) predicts unidirectional invisibility of microwave propagation at ZDCs through the open cavity magnonic system. Together with Eq.~(\ref{Symmetry}b), it implies a nonreciprocity that is mirror symmetric with respect to the cavity mode frequency. We highlight such striking effects in Figs.~\ref{fig:1}(b) and (c). In the following, we experimentally verify them.

\textit{Experimental results.-}First, we present results measured by placing the YIG sphere at the position with a balanced coupling ($J = \Gamma$). We focus on the cavity mode $\omega_{\rm{c}}/2\pi=4.724$ GHz. Figs.~\ref{fig:2}(a) and (b) show respectively the mapping of $|S_{21}(\omega)|$ and $|S_{12}(\omega)|$, which are plotted as a function of the frequency detuning ($\Delta_{\rm{c}} = \omega-\omega_{\rm{c}}$) and field detuning ($\Delta_{\rm{m}}=\omega_{\rm{m}}-\omega_{\rm{c}}$). The intrinsic damping rates of two hybridized modes are fitted from $|S_{21}(\omega)|$ and $|S_{12}(\omega)|$ spectra, and plotted as green circles in Figs.~\ref{fig:2}(c) and (d), respectively.  Theoretical eigenvalues, calculated from Eq.~(\ref{Hamiltonian}) using $J/2\pi=\Gamma/2\pi$ = 7.9 MHz, $\alpha/2\pi$ = 1.1 MHz and $\beta/2\pi$ = 15 MHz, are plotted for comparison~\cite{supp}. The real part eigenvalues Re($\widetilde{\omega}_{\pm}$), plotted as the black dashed lines in Figs.~\ref{fig:2}(a) and (b), agree well with the measured dispersion. The imaginary part eigenvalues $|$Im($\widetilde{\omega}_{\pm}$)$|$, plotted as black solid curves in Figs.~\ref{fig:2}(c) and (d), agree well with the measured damping rates. Two pairs of ZDCs predicted by Eq.~(\ref{Hamiltonian}) are observed, as marked by the arrows in Figs.~\ref{fig:2}(c) and (d).  

Unidirectional invisibility is also confirmed. Choosing the side ZDC marked by the blue arrow as an example, Figs.~\ref{fig:2}(e) and (f) show the measured $|S_{21}(\omega)|$ and $|S_{12}(\omega)|$ spectra, respectively. An ultrasharp dip in $|S_{21}(\omega)|$ is observed at $\omega = \omega_{-}$ = 4.615 GHz. The calculated result using $\kappa/2\pi$ = 880 MHz and $\gamma/2\pi$ = 0.071 MHz, based on fitting Eq.~(\ref{J-s12}) to the measured spectrum ~\cite{supp}, is plotted as the thin curves for comparison. These spectra show that microwave transmission from port 1 to port 2 is completely blocked at this frequency. Similarly, unidirectional invisibility is found at $\omega = \omega_{+}$ = 4.833 GHz, as shown by the $|S_{21}(\omega)|$ and $|S_{12}(\omega)|$ spectra plotted in Figs.~\ref{fig:2}(g) and (h), respectively, which are taken at the ZDC marked by the magenta arrow. Together, as highlighted in Figs.~\ref{fig:1}(b) and (c), results of Figs.~\ref{fig:2}(e) - (h) verify the mirror symmetric nonreciprocity that is summarized by Eq.~(\ref{Symmetry}). 

\begin{figure}[t]
  \centering
  \includegraphics[width=0.47\textwidth]{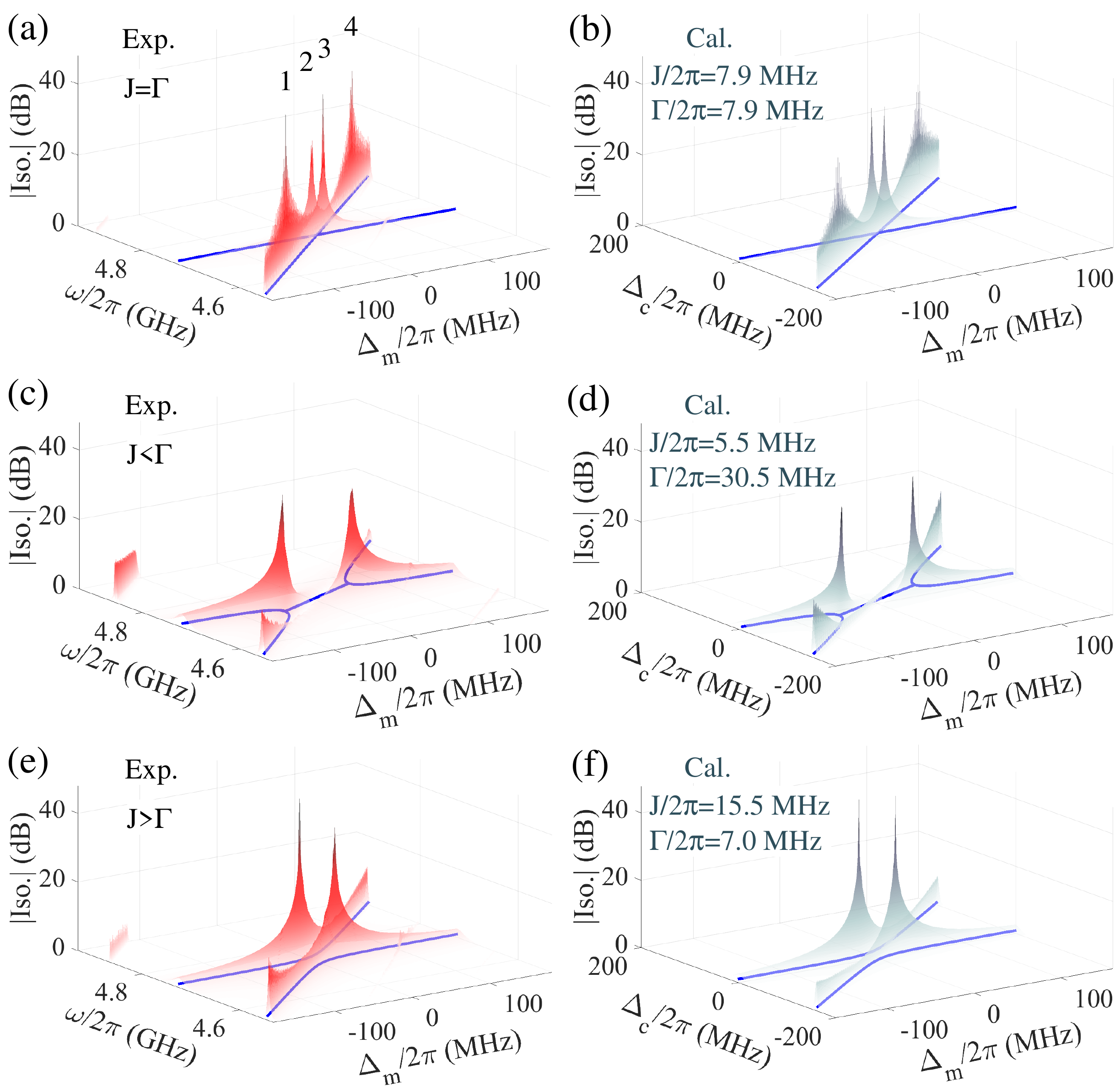}
  \caption{(color online). (a)(c)(e)~Measured isolation ratio as a function of the working frequency $\omega$ and the field detuning $\Delta_{\rm{m}}$. (a) $J=\Gamma$, (b) $J<\Gamma$, and (c) $J>\Gamma$. Four isolation peaks are labeled in (a). The blue solid curves are the dispersion of the coupled magnon-photon modes. (b)(d)(f)~Calculated isolation ratio with different coupling strengths.}
  \label{fig:3}
\end{figure}

Next, we study the nonreciprocity with different combinations of coherent and dissipative coupling rates. The difference between the forward ($S_{21}$) and backward ($S_{12}$) transmission amplitudes is extracted in dB scale (defined as 20$*$log$_{10}|S_{21}/S_{12}|$), and we take its absolute value as the isolation ratio (Iso.). For the situation of $J/2\pi=\Gamma/2\pi$ = 7.9 MHz discussed in Fig.~\ref{fig:2}, the measured isolation ratio as a function of the working frequency $\omega$ and the field detuning $\Delta_{\rm{m}}$ is plotted in Fig.~\ref{fig:3}(a). Four peaks appear at the two pairs of ZDCs, which we classify as central isolation peaks (2, 3) and side isolation peaks (1, 4). Away from the ZDCs the isolation ratio gradually decreases following the dispersion of the hybridized modes. The theoretical calculation shown in Fig.~\ref{fig:3}(b) is in good agreement with the experimental result. Here, at balanced coupling, the dispersion plotted in solid curves indicates that the magnon and photon modes are simply crossing. This is consistent with our previous study \cite{Harder-18}. Changing the position of the YIG sphere to adjust the coupling strengths, we find that the competition between coherent and dissipative couplings shapes the appearance of the isolation ratio.  When $J<\Gamma$, the pattern of the isolation ratio reveals level attraction as shown in Fig.~\ref{fig:3}(c). In the $J>\Gamma$ case, it shows level repulsion [Fig.~\ref{fig:3}(e)]. In both cases, the calculated results shown in Figs.~\ref{fig:3}(d) and (f) agree well with the measured data. The small isolation peak above 4.8 GHz in Fig.~\ref{fig:3}(c) and (e) is attributed to the effect of coupling between the cavity mode and a higher-order magnetostatic mode \cite{Dany-19,Harder-18s}, which we ignore in the calculation.

Apparently, this scheme of isolation is fundamentally different from conventional nonreciprocal response utilizing either Faraday rotation or ferromagnetic resonance \cite{Faraday,Hogan-53, Rowen-53,Adam-2002,Camley-09}, where large ferrites and complex port design (involving resistive sheets, quarter-wave plates, etc.) are needed, making it bulky for integration. In our case, the YIG sphere is only 1-mm in diameter. The isolation is the result of a synergetic contribution of the cavity mode, the magnon mode, and the interference between their coherent and dissipative interactions. The distinction is most clearly seen from the fact that the nonreciprocity disappears when either $J=0$ or $\Gamma=0$~\cite{supp}.

\begin{figure}[h]
	\centering
	\includegraphics[width=0.47\textwidth]{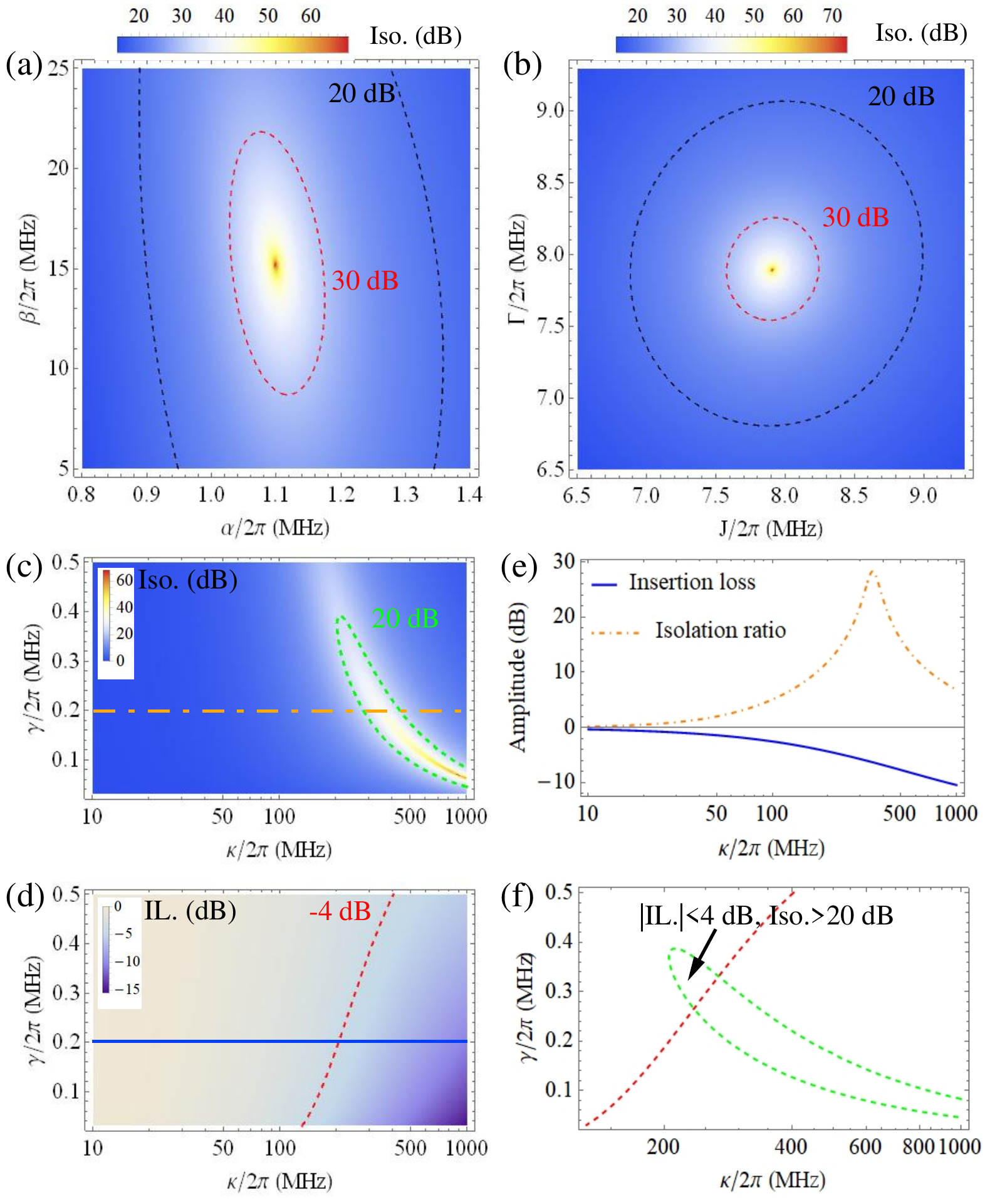}
	\caption{(color online). Isolation ratio near the side isolation peak plot as a function of (a) the intrinsic mode damping rates $\alpha$ and $\beta$, and (b) the coupling strengths $J$ and $\Gamma$. The black and red dashed lines are the contour lines of 20 dB and 30 dB isolation ratio, respectively. (c)~Isolation ratio as a function of the external damping rates $\kappa$ and $\gamma$. The green dashed line is the contour line of the isolation ratio equal to 20 dB. (d)~Insertion loss as a function of  $\kappa$ and $\gamma$. The red dashed line is the contour line of the insertion loss equal to -4 dB.  (e)~Amplitude of the isolation ratio and insertion loss as a function of $\kappa$ when $\gamma/2\pi=0.2$ MHz.  (f)~Contour lines extracted from (c) and (d). In the region marked with an arrow, the insertion loss is less than 4 dB, and the isolation ratio is larger than 20 dB.}
	\label{fig:4}
\end{figure}

Such unique nonreciprocal dynamics enable flexible controllability. In addition to the detuning that we have shown, the coupled system has a series of other control parameters. Taking a working frequency of $\omega/2\pi$ = 4.620 GHz as an example, Fig.~\ref{fig:4}(a) shows the isolation ratio near the side peak as a function of intrinsic damping rates $\alpha$ and $\beta$, which is calculated by fixing the coupling strengths $J/2\pi=\Gamma/2\pi$ = 7.9 MHz. The black and red dashed lines are the contour lines of 20 dB and 30 dB of isolation, respectively. In this case, the isolation ratio is more sensitive to $\alpha$ than $\beta$, because the side isolation is governed by the magnon-like hybridized mode. Similarly, isolation can be controlled by tuning the coupling rates $J$ and $\Gamma$, as shown in Fig.~\ref{fig:4}(b) by fixing $\alpha/2\pi=1.1$ MHz and $\beta/2\pi=15$ MHz. In this case, again, an isolation ratio above 30 dB can be achieved by adjusting the YIG position to control the coupling rates.

Finally, we note that for practical applications, another crucial performance index is the insertion loss (IL.). The most desirable performance is achieved with a high isolation ratio and low insertion loss, which can be optimized in our system by choosing suitable external damping rates $\kappa$ and $\gamma$, for the cavity and magnon modes, respectively. Setting $\alpha/2\pi=1.1$ MHz, $\beta/2\pi=15$ MHz, and $J/2\pi=7.9$ MHz, the isolation ratio and insertion loss as a function of $\kappa$ and $\gamma$ are shown in Figs.~\ref{fig:4}(c) and (d), respectively. The green dashed curve in Fig.~\ref{fig:4}(c) is the contour curve for 20 dB of isolation ratio and the red dashed curve in Fig.~\ref{fig:4}(d) is the contour curve for -4 dB of insertion loss. The horizontal lines in Figs.~\ref{fig:4}(c) and (d) are re-plotted in Fig.~\ref{fig:4}(e), showing the general trade-off between isolation ratio and insertion loss. The two contour curves in Figs.~\ref{fig:4}(c) and (d) are re-plotted in Fig.~\ref{fig:4}(f), where we can clearly find an optimized region with small insertion loss ($|$IL.$|<4~$dB) and considerably large isolation ratio ($|$Iso.$|>20~$dB) simultaneously. 

\textit{Conclusion.-}By designing an open cavity magnonic system to harness the cooperative effect of coherent and dissipative magnon-photon couplings, we demonstrate a new scheme for realizing nonreciprocal microwave transmission in the linear response regime. Zero-damping conditions for hybridized modes are realized, at which unidirectional invisibility of microwave propagation is discovered. These effects are well explained by a model that takes into account the direction-dependent relative phase between coherent and dissipative magnon-photon couplings, which breaks the time-reversal symmetry for microwave propagation. These nonreciprocal dynamics have flexible controllability, and enable optimized performance with large nonreciprocity and low insertion loss. The key physics of the interference between coherent and dissipative couplings is general, and may be applied to other systems without any external magnetic field, such as superconducting circuits in which coherent and dissipative couplings may be implemented and controlled by coupling the systems to the environment.\\

\begin{acknowledgments}

This work has been funded by NSERC Discovery Grants and NSERC Discovery Accelerator Supplements (C.-M. H.). J.Q.Y. is supported by the National Key Research and Development Program of China (Grant No. 2016YFA0301200) and the NSFC (Grant No. U1801661). We would like to thank T. Yu, J. Xiao, I. Proskurin, R. L. Stamps, P. Blunden, and M. Harder for discussions.

\end{acknowledgments}

\appendix

\end{document}